\documentclass[
  aps,
  10pt,
  physrev,
  superscriptaddress,
  amsmath,
  amssymb,
  amsfonts,
  floatfix,
  twocolumn,
  longbibliography
]{revtex4-2}

\usepackage{graphicx}
\usepackage{subfigure}
\usepackage{adjustbox}
\usepackage{bm}
\usepackage{color}
\usepackage{braket}
\usepackage{standalone}
\usepackage{multirow}
\usepackage{tikz}
\usepackage[utf8]{inputenc}
\usepackage{dsfont}
\usepackage[
  colorlinks,
  bookmarks=true,
  citecolor=blue,
  linkcolor=blue,
  urlcolor=blue
]{hyperref}
\usepackage{csquotes}

\usepackage{microtype}

\newcommand{\dd}[1]{\mathrm{d}#1\,}
\definecolor{dgreen}{rgb}{0.1,0.5,0.1}
\definecolor{babyblue}{rgb}{0.54, 0.81, 0.94}
\DeclareMathOperator{\tr}{tr}
\DeclareMathOperator{\e}{e}
\newcommand{\iu}{\mathrm{i}\mkern1mu}
\DeclareMathOperator{\diag}{diag}
\usepackage{nicefrac}

\begin{document}

\title{Quantum Non-Hermitian Topological Sensors}
\author{Florian Koch}
\affiliation{Institute of Theoretical Physics, Technische Universität Dresden and Würzburg-Dresden Cluster of Excellence ct.qmat, 01062 Dresden, Germany}
\author{Jan Carl Budich}
\email{jan.budich@tu-dresden.de}
\affiliation{Institute of Theoretical Physics, Technische Universität Dresden and Würzburg-Dresden Cluster of Excellence ct.qmat, 01062 Dresden, Germany}
\date{\today}

\begin{abstract}
  We investigate in the framework of quantum noise theory how the striking boundary-sensitivity recently discovered in the context of non-Hermitian (NH) topological phases may be harnessed to devise novel quantum sensors.
  Specifically, we study a quantum-optical setting of coupled modes arranged in an array with broken ring geometry that would realize a NH topological phase in the classical limit.
  Using methods from quantum-information theory of Gaussian states, we show that a small coupling induced between the ends of the broken ring may be detected with a precision that increases exponentially in the number of coupled modes, e.g.\ by heterodyne detection of two output modes.
  While this robust effect only relies on reaching a NH topological regime, we identify a resonance phenomenon without direct classical counterpart that provides an experimental knob for drastically enhancing the aforementioned exponential growth.
  Our findings pave the way towards designing quantum NH topological sensors (QUANTOS) that may observe with high precision any physical observable that couples to the boundary conditions of the device. \\
  \\
  \noindent DOI: \href{https://doi.org/10.1103/PhysRevResearch.4.013113}{10.1103/PhysRevResearch.4.013113}
\end{abstract}

\maketitle

\section{Introduction}
\begin{figure}[htp]
	\centering
	\includegraphics{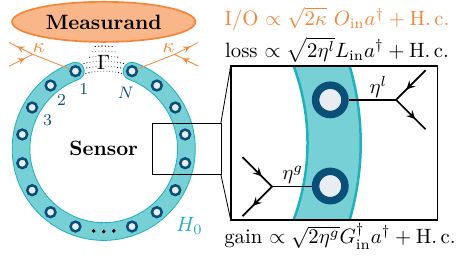}
	\caption{
    Illustration of the quantum non-Hermitian topological sensor (QUANTOS) setup.
    The device consists of an odd number of Bosonic modes (viewed as sites) arranged in an open ring geometry.
    The observed quantity (measurand) affects the weak coupling $\Gamma$ between the ends.
    Observed input/output ports with coupling rates $\kappa$ are indicated in orange.
    Dissipative quantum channels inducing a staggered pattern of gain (coupling $\eta^g$) and loss ($\eta^l$) are exemplified in black (see magnified inset).
    Formulas describe the underlying system-bath coupling Hamiltonians \cite{Gardiner2004,Gerry2005} of all channels. \vspace{-2mm}
  }
  \label{fig1}
\end{figure}

The quest for novel sensors that push the fundamental quantum-mechanical precision-limits represents a promising direction towards widely applicable quantum technology \cite{Degen2017,Acin2018,Pirandola2018}.
The basic physical mechanism underlying many quantum sensors may be understood as a detectable energy-level shift in response to an external perturbation \cite{Degen2017}.
In closed systems described by a Hermitian Hamiltonian, such energy shifts are always continuous towards perturbations, which may limit the achievable sensitivity of a given setting.
By contrast, the spectra of non-Hermitian (NH) Hamiltonians effectively describing dissipative systems \cite{Rotter2009,Moiseyev2011,Ashida2020} may exhibit non-analytic \cite{Berry2004,Heiss2012,Miri2019} and asymptotically even discontinuous behavior \cite{Reichel1992,Gong2018,Kunst2018,Yao2018,Herviou2019,Koch2020}, which in principle enables an unlimited spectral sensitivity.
Based on these insights, various architectures for NH sensors have been proposed \cite{Wiersig2014,Hodaei2017,Chen2017,Langbein2018,Lau2018,Zhang2019,Wang2020,Budich2020,McDonald2020,Wiersig2020}, some of which have already been experimentally realized \cite{Hodaei2017, Chen2017}.
Interestingly, combining NH sensing with the notion of topological matter \cite{Hasan2010,Qi2011,Wen2017,Bergholtz2021}, an enhancement in sensitivity that scales exponentially with system size is promoted to a stable phenomenon independent of fine-tuning \cite{Budich2020}.
More specifically, such non-Hermitian topological sensors are based on the energy-shift of a topological edge mode in response to small changes in the boundary conditions \cite{Gong2018,Kunst2018,Koch2020,Budich2020} of a chain in broken ring geometry (cf.~Fig.~\ref{fig1}).

Here, we develop a microscopic theory of quantum non-Hermitian topological sensors (QUANTOS) (see Fig.~\ref{fig1} for an illustration), that generalizes the aforementioned classical devices \cite{Budich2020} obtained in the limit of neglecting the quantum-mechanically inevitable input noise \cite{Gardiner2004,Clerk2010} to a fully quantum-mechanical setting in the framework of quantum Langevin equations \cite{Haken1970, Gardiner2004,Zhang2019,McDHanCle2021}.
Quite remarkably, we show that the classically derived exponential enhancement of sensitivity carries over to the noise-limited precision of a generic quantum-optical QUANTOS architecture.
To this end, we study the input-output relations associated with concrete experimental observation schemes such as heterodyne detection.
There, the probability density $p_\Gamma(x)$ for observing the output $x$ parametrically depends on the all-important boundary condition parameter $\Gamma$ that couples to the observable detected by the QUANTOS (see Fig.~\ref{fig1}).
The uncertainty $\Delta \Gamma$ in assessing the value of $\Gamma$ is then limited by the Fisher information $\mathcal I [p_\Gamma]$ \cite{Fisher1922} via the Cramer-Rao bound \cite{Rao1945,Cramer1946,Kay1993}
\begin{align}
\Delta \Gamma \ge \mathcal I^{-\frac{1}{2}} = \kappa \exp(-\alpha N).
\label{eqn:one}
\end{align}
The latter equality in Eq.~(\ref{eqn:one}), where $N$ denotes the number of modes (system size) of the QUANTOS and $\kappa, \alpha >0$, may be seen as a quantum mechanical counterpart of the exponentially enhanced sensitivity reported in a classical context in Ref.~\cite{Budich2020}.
Given the fact that the Cramer-Rao bound, i.e. the first relation in Eq.~(\ref{eqn:one}) can in principle be saturated \cite{Kay1993,Zhang2019}, our main finding is that the quantum noise-limited precision of the QUANTOS is indeed exponentially enhanced with $N$ in a wide parameter range.
Furthermore, we reveal a resonance phenomenon that provides an experimental knob for tuning the value of $\alpha$ (see Eq.~(\ref{eqn:one})) so as to drastically increase the exponential growth rate of the precision.\\

\section{Non-Hermitian topology by quantum dissipation}
We now set up a dissipative quantum-optical framework of coupled modes which yields a NH topological tight-binding model in the classical limit of neglecting the input noise.
To this end, we consider a vector $a = (a_1(t),\ldots, a_{N}(t))^T$ of Bosonic modes oscillating with an optical frequency $\omega_0$, the weakly coupled (as compared to $\omega_0$) \cite{Joshi2014} dynamics of which is governed by the quantum Langevin equation ~\cite{Gardiner2004, Zhang2019} (see also Appendix \ref{app:diss_multi}) ($\hbar =1$)
\begin{align}
\iu\partial_t \, a = (\mathcal H_0 + \iu \eta) \, a + \iu F = \tilde{\mathcal H} \, a + \iu F,
\label{eqn:qL}
\end{align}
where $\mathcal H_0$ is the Hermitian part describing the coherent coupling between the modes in a rotating frame with respect to $\omega_0$, which, together with the anti-Hermitian overall gain ($\eta_j >0$) or loss ($\eta_j<0$) rates $\eta = \text{diag}(\eta_1, \ldots, \eta_N)$ forms the classical NH Hamiltonian $\tilde{\mathcal H}$ \footnote{Note that we have assumed a mode basis in which the anti-Hermitian part $i \eta$ of $\tilde{\mathcal H}$ is diagonal, corresponding to local dissipation of modes. Note that, up to a unitary basis transformation, this setting is sufficient to realize any NH Hamiltonian $\tilde{\mathcal H}$ \cite{som}.}.
The input noise term $F = (F_1(t),\ldots, F_N(t))$ accounts for quantum fluctuations due to the coupling of the system to various dissipative channels detailed in the following. Specifically, we consider two different levels of dissipation: First, observed channels that are assumed to be controlled by an observer probing the input-output relations of the QUANTOS device. Second, unobserved channels that introduce optical loss and gain, respectively.
This leads to the decomposition (cf.~Eq.~(\ref{eqn:qL}))
\begin{align}
&\eta_j = - \kappa_j - \eta^l_j + \eta^g_j, \nonumber \\
&F_j = \sqrt{2\kappa_j}  O_{j,\text{in}} + \sqrt{2\eta^l_j} L_{j,\text{in}} - \sqrt{2\eta^g_j} G_{j,\text{in}}^\dag,
\label{eqn:noiseterms}
\end{align}
where the coupling rates $\kappa_j \ge 0$ and input noise operators $O_{j,\text{in}}(t)$ model an observed channel for mode $a_j$.
Similarly, $\eta^l_j\ge 0$ along with $L_{j,\text{in}}$ correspond to unobserved loss channels and $\eta^g_j\ge 0$ along with $G_{j,\text{in}}^\dag$ to unobserved gain channels, respectively.

\begin{figure}[htp]
	\centering
	\includegraphics{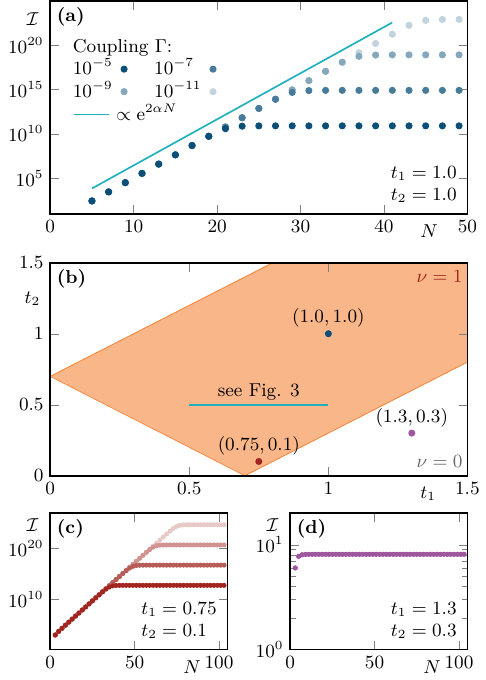}
	\caption{
    (a) Scaling with system size $N$ of the Fisher information $\mathcal I$ corresponding to a Heterodyne detection scheme for several values of $\Gamma$ (see plot legends).
    The exponential law corresponding to Eq.~(\ref{eqn:one}) with $2\alpha = 0.43011$ is shown as a guide to the eye.
    (b) Topological phase diagram (defined by integer invariant $\nu$) of the model (\ref{eqn:hnh}) as a function of parameters $t_1>0$ and $t_2>0$.
    Other parameter-regimes are readily inferred using the symmetries $\nu(-t_1,t_2) = -\nu(t_1,t_2)$ and $\nu(t_1,-t_2) =\nu(t_1,t_2)$.
    Parameter points used in the three other panels as well as in Fig.~\ref{fig:resonance} are indicated.
    (c) $\mathcal I(N)$ for another parameter set in the topological region ($\nu =1$) confirms qualitative similarity to panel (a).
    (d) $\mathcal I(N)$ for a parameter set in the topologically trivial phase ($\nu =0$) does not exhibit the exponential increase with $N$ that is characteristic of the QUANTOS.
    Other model parameters are $\gamma = 0.7, \omega = 0$ in all plots.
  }
	\label{fig:exp_amp_region}
\end{figure}

To model a basic QUANTOS device, we aim at bringing $\tilde{\mathcal H}$ (see Eq.~(\ref{eqn:qL})) into a NH topological phase.
To this end, we consider a staggered gain-loss pattern with $\eta^l_{2n-1} = \eta^g_{2n} = \gamma$ and zero couplings otherwise.
This structure suggests the definition of a unit-cell (indexed by $n$) that consists of the two modes $a_{2n-1}$ and $a_{2n}$ and to consider the NH Bloch band structure of a system that is translation invariant with respect to this unit cell.
In this language, our gain-loss pattern results in a term $-\iu \gamma\sigma_z$ in $\tilde{\mathcal H}$, where $\sigma_j,~j=x,y,z$ denote the standard Pauli matrices acting within the two-mode unit cell.
Regarding the Hermitian part $\mathcal{H}_0$, we consider a mode-coupling between nearest-neighbor unit cells that gives rise to the Bloch Hamiltonian $H_0(k) = (t_1 + t_2 \cos(k))\sigma_x + t_2 \sin(k) \sigma_z$.
Overall, the effective NH Bloch Hamiltonian then reads as
\begin{align}
\tilde H(k) =  (t_1 + t_2 \cos(k))\sigma_x + (t_2 \sin(k) - \iu\gamma) \sigma_z.
\label{eqn:hnh}
\end{align}
The NH topological phase of $H$ is determined by the integer quantized spectral winding number $\nu = \frac{1}{2\pi \iu}\oint_0^{2\pi} \dd{k} \{\partial_k\ln[\det(\tilde H(k))]\}$ \cite{Gong2018,Bergholtz2021}.
As for their classical counterparts, the precision of the QUANTOS device is found to crucially rely on a non-zero value of $\nu$, i.e.~on reaching a topologically non-trivial NH phase (see Fig.~\ref{fig:exp_amp_region}(b) for the topological phase diagram of the NH Hamiltonian (\ref{eqn:hnh})).
Note that $\tilde{\mathcal H}$ fulfills a local parity-time ($PT$) symmetry that is known to entail possible inconsistencies \cite{Lee2014,Tang2016} when considering  $PT$-symmetric quantum mechanics as a fundamental theory \cite{Bender1998}.
However, here this symmetry is only present due to the simplifying parameter choice of balanced gain and loss.
Breaking this coincidental symmetry by slightly unbalancing gain and loss has no substantial effect on the exponential
enhancement (cf.~Eq.~(\ref{eqn:one}) and Appendix \ref{app:balanced}).

Regarding the observed channels probing the output of the system, we generally consider $\kappa_1 = \kappa_N =\gamma >0$ and $\kappa_j =0,~j\ne 1,N$ in the following (see orange channels in Fig.~\ref{fig1}).
This choice is motivated by the aforementioned crucial role that the level shift of the topological edge mode plays in the working principle of the QUANTOS, i.e.~the input-output ports of the device should have a significant overlap with that boundary mode.
Regarding the unobserved channels, we assume vacuum input fields corresponding to vanishing expectation values $\langle L_{j,\text{in}}\rangle = \langle G_{j,\text{in}}^\dag \rangle =0$ of the reservoir field operators.
Furthermore, all channels are assumed to be mutually independent and Markovian (local white-noise limit) \cite{Joshi2014}.
Finally, the number of sites $N$ is chosen to be odd so as to stabilize a single zero-energy edge mode in the case of open boundary conditions ($\Gamma = 0$) \cite{Kunst2018,Budich2020}.
The coupling $\Gamma$ between the ends of the ring that is modulated by the measurand (see Fig.~\ref{fig1}) is modeled by the term $\Gamma (a_N^\dag a_1 +  a_1^\dag a_N)$ entering $\mathcal H_0$ which may be thought of as a generic tunnel coupling between the ends.

\section{Input-Output Theory of QUANTOS}
The observed channels satisfy the standard input-output relation (cf.~\cite{Gardiner2004})
\begin{align}
O_{j, \text{out}} = O_{j, \text{in}} - \sqrt{2\kappa_j} a_j.
\label{eqn:io}
\end{align}

Our goal is to express the system operators $a_j$ in terms of the input operators $O_{j, \text{in}}$ by means of the Green's function $\mathcal{G} = (\partial_t + i \tilde{\mathcal H})^{-1}$ that accounts for the (dissipative) propagation of the input signal through the system.
To this end, it is convenient to transform the quantum Langevin equation (\ref{eqn:qL}) into frequency space and to introduce the quadrature operators $q = a + a^\dagger$ as well as $p = \iu(a^\dagger - a)$ for the the system modes, and similarly $Q,P$ for the observed, and $Q',P'$ for the unobserved input operators.
Then, the quantum Langevin equation (\ref{eqn:qL})  becomes an algebraic equation which is solved in terms of the Green's function so as to eliminate the occurrence of the system operators $a_j$ from the input-output relation (\ref{eqn:io}).
Explicitly, we obtain \cite{Zhang2019}
\begin{align}
   \begin{pmatrix}
       Q_\text{out}[\omega] \\
       P_\text{out}[\omega]
   \end{pmatrix} = S[\omega]
   \begin{pmatrix}
       Q_\text{in}[\omega] \\
       P_\text{in}[\omega]
   \end{pmatrix} + L[\omega]
   \begin{pmatrix}
       Q'_\text{in}[\omega] \\
       Q'_\text{in}[-\omega] \\
       P'_\text{in}[\omega] \\
       P'_\text{in}[-\omega],
   \end{pmatrix}
\label{eqn:IO}
\end{align}
where we have defined the scattering matrix $S[\omega] = 1 - K_\text{o}^T \tilde{ \mathcal{G}}[\omega] K_o$ as well as the noise matrix $L = -K_o^T \tilde{ \mathcal{G}}[\omega] K_u$ using the frequency space Green's function $\tilde{\mathcal{G}}[\omega]$ in the quadrature basis.
In our framework, the coupling matrix of the observed channels $K_o$ is a diagonal $2N \times 2N$-matrix with non-vanishing elements $K_o^{j\ j} = K_o^{(N+j)\ (N+j)} = \sqrt{2\gamma}$ only for the observed sites $j=1,N$.
Similarly, the coupling matrix $K_u$ of the unobserved channels is a $2N \times 4N$-matrix with $K_u^{j\ j} = K_u^{(N+j)\ (2N+j)} = \sqrt{2\gamma}$ if $j$ is a site with loss (odd $j$ in our setting), while $K_u^{j\ (N+j)} = -K_u^{(2N + j)\ (3N+j)} = -\sqrt{2\gamma}$ if site $j$ exhibits gain (even $j$ in our setting).
Using Eq.~(\ref{eqn:IO}), we may compute the response of the system to an arbitrary input.

Accounting for the linear structure of our model system, it is natural to consider Gaussian input states which are fully described in the quadrature basis by their amplitude vector $\mu$ and their covariance matrix $V$, respectively \cite{Weedbrook2012}.
In this framework, basically any observable or correlation function may be readily computed in a numerically exact fashion, noting that the Gaussian character of the input states is preserved by the considered scattering dynamics.
Specifically, the relevant input-output relations then read as (cf. Eq.~(\ref{eqn:IO}))
\begin{align}
    \mu_\text{out} = S \mu_\text{in},\quad V_\text{out} = S V_\text{in} S^T + L V'_\text{in} L^T.
    \label{eqn:GaussianIO}
\end{align}

\begin{figure}[htp]
	\centering
	\includegraphics{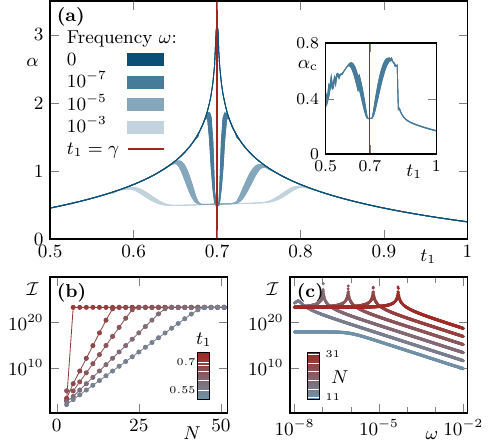}
	\caption{
    (a) Dependence of the exponential growth rate $\alpha$ (cf. Eq.~(\ref{eqn:one})) as a function of $t_1$ at $t_2=0.5$ for several values of frequency $\omega$ (relative to the free mode frequency $\omega_0$).
    A sharp resonance around $\gamma = t_1$ is visible at $\omega = 0$ that is pinched off at a finite maximum value of $\alpha$ for finite frequency shifts.
    Inset: Corresponding exponential growth rate $\alpha_c$ of the energy level shift known from the classical NH Hamiltonian theory.
    (b) Steepening of the exponential growth of $\mathcal I (N)$ when approaching the resonance at $\gamma = t_1 =0.7, \omega =0$ by varying $t_1$.
    (c) Frequency-dependence of $\mathcal I$ at various system sizes $N$ at $t_1 =0.69, t_2 =0.5$, exhibiting a striking $N$-dependent frequency-shifted resonance.
    Other parameters are $\gamma =0.7, \Gamma = 10^{-11}$ in all plots.
  }
	\label{fig:resonance}
\end{figure}

\section{Fisher information and Cramer-Rao bound}
Equipped with a constructive recipe for computing the exact input-output relations of the QUANTOS, we now turn to assessing the precision of the device.
This amounts to systematically estimating the quantum noise-limited precision with which the boundary condition parameter $\Gamma$ (see Fig.~\ref{fig1}) can be determined.
Generally speaking, the quantum mechanical probability for observing a measurement outcome $x$ in a given experimental setting will follow a probability distribution $p_\Gamma(x)$ that may parametrically depend on $\Gamma$.
The sensitivity of this distribution to small changes in $\Gamma$ is quantified by the Fisher information $\mathcal{I} = \int \dd{x}\left\{\partial_\Gamma \ln[ p_\Gamma(x)]\right\}^2$ \cite{Fisher1922,Kay1993}.
Via the Cramer-Rao bound $\Delta \Gamma \ge \frac{1}{\sqrt{\mathcal I}}$ (cf.~Eq.~(\ref{eqn:one})), the Fisher-information is directly linked to the uncertainty (or variance) $\Delta \Gamma$ with which $\Gamma$ may be estimated using the given experimental scheme.
In this context, it is important to note that the Cramer-Rao bound can, at least in principle, be saturated \cite{Kay1993,Zhang2019} so as to achieve the optimal precision $\Delta \Gamma = \frac{1}{\sqrt{\mathcal I}}$.
We note that, beyond our present context of Cramer-Rao bounds, the (quantum) Fisher information has found wide applications in physics including the derivation of quantum speed limits \cite{LiebRobinson}, where it has recently been connected to the presence of geometric phases \cite{SunZhe2019}.

In our present context of multivariate Gaussian distributions, where both the mean vector $\mu$ and the covariance matrix $V$ depend on the parameter $\Gamma$, the Fisher information is given by \cite{Kay1993, Oh2019}
\begin{align}
    \mathcal{I} = \frac{1}{2}\tr\left[V^{-1} \frac{\partial V}{\partial \Gamma} V^{-1} \frac{\partial V}{\partial \Gamma}\right] + \left(\frac{\partial \mu}{\partial \Gamma}\right)^T V^{-1} \left(\frac{\partial \mu}{\partial \Gamma}\right).
    \label{eqn:GaussFisher}
\end{align}

For concreteness, we illustrate our results in the following for a heterodyne detection scheme.
If we measure all observed channels, the heterodyne detection yields a multivariate Gaussian probability density where the mean vector is equal to the amplitude vector $\mu_\text{out}$ and the covariance matrix is equal to $V_\text{out} + 1$ (cf.~Eq.~(\ref{eqn:GaussianIO})).
This amounts to inserting  $\mu = \mu_\text{out}$ and $V = V_\text{out} + 1$ into Eq.~(\ref{eqn:GaussFisher}), where the added noise in the covariance matrix stems from the simultaneous detection of the canonically conjugated $Q$ and $P$ quadratures that is characteristic of heterodyne detection.

\section{Exponentially enhanced precision}
In Fig.~\ref{fig:exp_amp_region}, we present numerical data on the Fisher information $\mathcal I$ (cf.~Eq.~(\ref{eqn:GaussFisher})) as a function of system size $N$.
For parameter values in the topologically non-trivial hallmarked by a non-vanishing value of the spectral winding number $\nu$ (see topological phase diagram in Fig.~\ref{fig:exp_amp_region}(b)) QUANTOS region, i.e.\ in panels (a) and (c) of Fig.~\ref{fig:exp_amp_region}, we find that $\mathcal I$ grows exponentially with $N$ (cf.\ Eq.~(\ref{eqn:one})) until it saturates towards a value that increases with decreasing $\Gamma$.
By contrast, in the topologically trivial parameter regime (see Fig.~\ref{fig:exp_amp_region}(d)), no systematic exponential growth of $\mathcal I (N)$ is observed.
Thus, while the precise value of $\alpha$ (cf.~Eq.~(\ref{eqn:one})) is parameter dependent, a finite positive $\alpha$ only exists for non-zero $\nu$.
In this sense, the observed enhanced sensitivity itself is a topologically stable phenomenon.
We identify this behavior and its dependence on the spectral winding number $\nu$ as the quantum-physical counterpart of the exponential energy level-shift $\Delta E_0 \sim \Gamma \exp(\alpha_c N)$ that the edge-state of the NH Hamiltonian $\tilde{\mathcal H}$ has been found to exhibit in the classical limit \cite{Budich2020}.
Via the Cramer-Rao bound (cf.~Eq.~(\ref{eqn:one})), this previously observed spectral sensitivity of $\tilde{\mathcal H}$ is thus found to generally carry over to an exponentially enhanced precision $\Delta \Gamma$ in quantum-noise limited measurements.

In addition to the qualitative dependence of $\mathcal I$ on the NH topological phase, we observe a striking resonance phenomenon within the $\nu =1$ QUANTOS regime that provides a knob for drastically increasing the exponential growth rate $\alpha$ (see Eq.~(\ref{eqn:one})), and that has no direct classical counterpart.
Numerical data on this intriguing behavior is presented in Fig.~\ref{fig:resonance}.
Specifically, in Fig.~\ref{fig:resonance}(a), we show the dependence of $\alpha$ on the hopping rate $t_1$.
We find evidence for a divergence (within the considered finite parameter values and numerical precision) of $\alpha$ at the parameter line $t_1 = \gamma$ for frequency $\omega =0$ in our rotating frame, i.e.\ at the free mode-frequency $\omega_0$ in the lab-frame.
For finite frequency detuning, this sharp divergence is cut off at a finite $\alpha$.
Quite remarkably, $t_1=\gamma$ is precisely the parameter-line where the individual unit cells of our model, independently described at $t_2=0$ by the local NH matrix $\tilde h_L = t_1 \sigma_x - \iu \gamma \sigma_z$ (cf.~Eq.~(\ref{eqn:hnh})) would exhibit exceptional points.
Yet, the aforementioned classical growth rate $\alpha_c$ of the energy-level shift $\Delta E_0$ \cite{Budich2020} stays bounded at this special parameter set (see inset of  Fig.~\ref{fig:resonance}(a)).
In Fig.~\ref{fig:resonance}(b), we illustrate how tuning towards the sweet-spot $t_1 = \gamma$ allows for a faster and faster exponential growth of $\mathcal I$, and thus for designing a high-precision QUANTOS with a fairly small number of coupled modes $N$.
Finally, in Fig.~\ref{fig:resonance}(c), we study the frequency-dependence of $\mathcal I$ for various choices of $N$ in a regime where the exponential growth with respect to $N$ is saturating.
That way, on top of the aforementioned behavior at fixed frequency, we observe a sharp resonance with respect to $\omega$, implying that $\mathcal I$ overshoots its saturated value by several orders of magnitude, if the probe signal of the QUANTOS is tuned towards the optimal value of $\omega$.

\section{Concluding discussion}
Inspired by the classical analysis of Ref.~\cite{Budich2020}, we have presented a quantum theory for a quantum non-Hermitian topological sensor (QUANTOS), the precision of which grows exponentially in system size $N$ in a wide parameter range, provided that the underlying coupled mode model is brought into a NH topological phase (cf.~Fig.~\ref{fig:exp_amp_region}).
The generic operating principle of this novel class of sensors naturally assumes an \enquote{off-state} in which the coupling $\Gamma$ between the ends of a broken ring geometry (cf.~Fig.~\ref{fig1}) are nearly uncoupled $\Gamma \simeq 0$.
In this regime, a weak link (finite change in $\Gamma$) induced by a measurand can be detected with high precision $\Delta \Gamma$ (cf.~Eq.~(\ref{eqn:one})).
This principle allows for a wide range of applications regarding the concrete choice of physical observables to be detected by QUANTOS devices, i.e.\ basically any entity that modifies the tunnel-coupling $\Gamma$ between the end modes qualifies as a measurand.
Complementary to our present approach of using an energy-level shift of a topological edge mode, in Ref.~\cite{McDonald2020} a local coupling between two non-reciprocal NH chains of opposite chirality, that are realized as conjugate quadratures of Bosonic modes, has been proposed as a candidate for exponentially enhanced quantum sensing. Different from our setting, active amplification by optical gain is not a necessary ingredient of that architecture, whereas a possible tradeoff lies in a certain susceptibility with respect to an undesirable local coupling between the quadratures.

While the relation between NH topological boundary modes and the occurrence of an exponential sensitivity has a classical counterpart \cite{Budich2020}, here we have not only generalized this intriguing effect to a fully quantum-mechanical level but also identified new experimental possibilities for optimizing the precision of the QUANTOS by exploiting resonance phenomena (cf.\ Fig.~\ref{fig:resonance}) that do not have a direct counterpart in the classical limit.
Based on the generic theoretical modeling and analysis presented in this work, exploring experimental implementation of QUANTOS devices by microscopically describing the coupling $\Gamma$ to a desired measurand is an interesting subject of future work.

\begin{acknowledgments}
  We would like to thank Emil Bergholtz and Hannes Pichler for discussions.
  We acknowledge financial support from the German Research Foundation (DFG) through the Collaborative Research Centre SFB 1143, the Cluster of Excellence  ct.qmat, and the DFG Project No.~459864239.
  Our numerical calculations were performed on resources at the TU Dresden   Center for Information Services and High Performance Computing (ZIH).
\end{acknowledgments}

\appendix
\section{Damping in a single mode system}
We start by considering a single mode $a$ which is coupled to a bath with modes $b_\Omega$.
The total system is described by the Hamiltonian
\begin{align}
  H &= H_\text{system} + H_\text{bath} + H_\text{interaction} \\
  H_\text{system} &= \omega_0 a^\dagger a \\
  H_\text{bath} &= \int_0^\infty \dd{\Omega} \Omega b_\Omega^\dagger b_\Omega \label{eq:H_bath}\\
  H_\text{interaction} &= \int_0^\infty \dd{\Omega} f(\Omega) \left(a + a^\dagger\right)\left(b_\Omega + b_\Omega^\dagger\right) \label{eq:H_interaction}
\end{align}
which can be simplified in terms of the rotating wave approximation.
Here, we neglect the rapidly oscillating terms $b_\Omega^\dagger a^\dagger$ and $b_\Omega a$. Further, we expand the integration range to $(-\infty,\ +\infty)$ and argue that the \enquote{added terms} are far from resonance and thus negligible.
Finally, Eqs. (\ref{eq:H_bath}) and (\ref{eq:H_interaction}) are replaced by
\begin{align}
  H_\text{bath} &= \int_{-\infty}^{+\infty} \dd{\Omega} \Omega b_\Omega^\dagger b_\Omega\\
  H_\text{interaction} &= \int_{-\infty}^{+\infty} \dd{\Omega} f(\Omega) \left(b_\Omega^\dagger a + a^\dagger b_\Omega\right).
\end{align}

From the Hamiltonian we can derive the equations of motion
\begin{align}
  \dot a &= \iu [H, a] = -\iu\omega_0 a - \iu\int_{-\infty}^{+\infty} \dd{\Omega} f(\Omega) b_\Omega, \\
  \dot b_\Omega &= \iu [H, b_\Omega] = -\iu \Omega b_\Omega -\iu f(\Omega) a.
\end{align}
Formally integrating the equation of the bath modes gives
\begin{align}
  b_\Omega(t) = b_\Omega(0) \e^{-\iu\Omega t} -\iu f(\Omega) \int_0^t \dd{t'} a(t')\e^{-\iu\Omega(t - t')}.
\end{align}
Inserting this into the equation of motion of the system mode we get a differential-integro equation
\begin{align}
  \dot a(t)
  %&= -i\omega_0 a(t) - i \sum_\Omega f(\Omega) \left(b_\Omega(0) \e^{-i\Omega t} -i f(\Omega) \int_0^t \text{d}t' a(t')\e^{-i\Omega(t - t')}\right) \\
  &= -\iu\omega_0 a(t) - \iu \int_{-\infty}^{+\infty} \dd{\Omega} f(\Omega) b_\Omega(0) \e^{-\iu\Omega t}  \nonumber \\
  &\qquad- \int_{-\infty}^{+\infty} \dd{\Omega} f(\Omega)^2 \int_0^t \dd{t'} a(t') \e^{-\iu\Omega(t-t')}. \label{eq:diff_int}
\end{align}

At this point we apply the \emph{Born-Markov approximations}.
Those approximations are used to describe systems with weak system-bath coupling (Born) in the limit of vanishing bath memory (Markov).
They state that only frequencies which are close to $\omega_0$ have an influence and terms that are further away are negligible.
Thus, we can approximate the coupling constant with its value at the mode-frequency $\omega_0$, i.e. $f(\Omega) \approx f(\omega_0) = \sqrt{\eta^l / \pi}$.
Inserting this into Eq.~(\ref{eq:diff_int}) allows to simplify the third term by
\begin{align}
  &- \int_{-\infty}^{+\infty} \dd{\Omega} f(\Omega)^2 \int_0^t \dd{t'} a(t') \e^{-\iu\Omega(t-t')} \\
  \approx &- \int_{-\infty}^{+\infty} \dd{\Omega} \frac{\eta^l}{\pi} \int_0^t \dd{t'} a(t') \e^{-\iu\Omega(t-t')} \\
  = &- \eta^l a(t) \label{eq:born_markov}
\end{align}
Further, we define
\begin{align}
  L_\text{in}(t)
  &= -\frac{\iu}{\sqrt{2\eta^l}}\int_{-\infty}^{+\infty} \dd{\Omega} f(\Omega) b_\Omega(0) \e^{-\iu\Omega t} \\
  &= -\frac{\iu}{\sqrt{2\pi}} \int_{-\infty}^{+\infty} \dd{\Omega} b_\Omega(0) \e^{-\iu\Omega t}. \label{eq:L_in}
\end{align}
Inserting Eqs. (\ref{eq:born_markov}) and (\ref{eq:L_in}) into Eq.~(\ref{eq:diff_int}) gives the quantum Langevin equation
\begin{align}
  \dot a = -\iu\omega_0 a(t) - \eta^l a(t) + \sqrt{2\eta^l} L_\text{in}(t).
\end{align}

Note that if the modes $b_\Omega$ are in a vacuum state, i.e.\ $\braket{0|b_\Omega|0} = 0$, we have $\braket{0|L_\text{in}(t)|0} = 0$.
Further, evaluation of the correlations show that the input noise is delta-normalized, i.e. $\braket{0|L_\text{in}(t) L_\text{in}^\dagger(t')|0} = \delta(t-t')$.
Thus, the so-called \emph{White noise limit} is fulfilled.

If we are interested in measuring the mode (observed channels), we repeat this calculation but integrate from $t \to \infty$ instead of $0 \to t$.
In order to distinguish observed and unobserved channels, we rename the noise operators $L \to O$. The calculation gives a similar result:
\begin{align}
  \dot a = -\iu\omega_0 a(t) + \eta^l a(t) + \sqrt{2\eta^l} O_\text{out}(t).
\end{align}
Subtracting the two differential equations of the input and output noise from each other gives the input-output-relation
\begin{align}
  O_\text{out}(t) = O_\text{in}(t) - \sqrt{2 \eta^l} a(t)
\end{align}
which can be used to model the measurement of mode $a(t)$ (cf. \cite{Gardiner2004}).

\section{Amplification of a single mode system}
In order to model gain of a single mode system, we once again couple the mode $a$ to a bath consisting of modes $h_\Omega$.
Here, the modes $h_\Omega$ describe an inverted oscillator heat bath.
After the use of the rotating wave approximation and the transformation $h_\Omega \to b_\Omega^\dagger$, the interaction Hamiltonian consists of two-mode squeezing terms \cite{Gerry2005, Zhang2019}
\begin{align}
  H_\text{interaction} = \int_{-\infty}^{+\infty} \dd{\Omega} f(\Omega) \left(b_\Omega^\dagger a^\dagger + a b_\Omega\right)
\end{align}

Thus, the equations of motion are
\begin{align}
  \dot a &= -\iu\omega_0 a(t) - \iu \int_{-\infty}^{+\infty} \dd{\Omega} f(\Omega) b_\Omega^\dagger, \\
  %\dot a^\dagger &= \iu\omega_0 a^\dagger(t) + \iu \int_{-\infty}^{+\infty} \dd{\Omega} f(\Omega) b_\Omega, \\
  %\dot b_\Omega &= -\iu\Omega b_\Omega - \iu f(\Omega) a^\dagger, \\
  \dot b_\Omega^\dagger &= \iu\Omega b_\Omega^\dagger + \iu f(\Omega) a.
\end{align}
Formally integrating the equations of motion for the bath modes $b_\Omega^\dagger$ leads to
\begin{align}
  b_\Omega^\dagger(t) = b_\Omega^\dagger(0) \e^{\iu\Omega t} + \iu f(\Omega)\int_0^t \dd{t'} a(t') \e^{\iu\Omega (t-t')}
\end{align}
which can be inserted into the equation of motion for $a$ and then be approximated within the aforementioned Born-Markov approximations by
\begin{align}
  \dot a
  %&= -i\omega_0 a(t) - i\sum_\Omega f(\Omega) \left(b_\Omega^\dagger(0) \e^{i\Omega t} + i f(\Omega)\int_0^t \text{d}t' a(t') \e^{i\Omega (t-t')}\right) \\
  &= -\iu \omega_0 a(t) - \iu \int_{-\infty}^{+\infty} \dd{\Omega} f(\Omega) b_\Omega^\dagger(0) \e^{\iu\Omega t} \nonumber \\
  &\qquad+ \int_{-\infty}^{+\infty} \dd{\Omega} f(\Omega)^2 \int_0^t \dd{t'} a(t') \e^{\iu\Omega (t-t')} \\
  &\approx -\iu \omega_0 a(t) - \frac{1}{\sqrt{2\eta^g}} G_\text{in}^\dagger(t) + \eta^g a(t).
\end{align}

Note that if the modes $b_\Omega$ are assumed to be in a vacuum state, i.e.\ $\braket{0|b_\Omega|0} =\braket{0|b^\dagger_\Omega|0} = 0$,  we obtain $\braket{0 | G_\text{in}^\dagger(t) | 0} = 0$.
Further evaluation of the correlations gives $\braket{0 | G_\text{in}(t) G_\text{in}^\dagger(t') |0} \approx \delta(t-t')$.

\section{Dissipation in a multi-mode system}
\label{app:diss_multi}
Next, we generalize our discussion to a multi-mode system described by $N$ modes $a_i,~i=1,\ldots,N$.
According to \cite{Joshi2014}, if we are in a weak coupling regime (weak as compared to the mode frequency $\omega_0$), a local description of the system-environment-interaction is valid.
That means that we may couple each mode $a_i$ locally to its individual bath $\sum_\Omega b_\Omega^\dagger b_\Omega$.
Thus we get a quantum Langevin equation as derived above for every mode $a_i$.
In addition, the weak coherent coupling between the modes is described by a Hermitian Hamiltonian $H_0$.
With these ingredients, we can construct a quantum theory corresponding to any given non-Hermitian Hamiltonian of the form $H = \sum_{ij} \bar{a}^\dagger_i \mathcal{H}_{ij} \bar{a}_j$.
To this end, we apply the following general procedure
\begin{enumerate}
  \item Split the non-Hermitian Hamiltonian matrix $\mathcal H$ into the Hermitian $\bar{\mathcal H}_0$ and the anti-Hermitian $\mathcal{H}_\text{AH}$ part:
  \begin{align}
    \mathcal{H} = \bar{\mathcal H}_0 + \mathcal{H}_\text{AH}.
  \end{align}
  The anti-hermitian part can be diagonalized with purely imaginary eigenvalues $\iu\eta_j\in i\mathbb{R}$.
  \item Switch to the basis in which the anti-Hermitian part is diagonal, i.e.\ $\bar a \to a$, by the unitary transformation:
  \begin{align}
    \tilde{\mathcal{H}} = U^\dagger \mathcal{H} U = U^\dagger \bar{\mathcal H}_0 U + \iu\diag(\eta_j).
  \end{align}
  \item Calculate the transformed equations of motion:
  \begin{align}
    \iu\partial_t a = \tilde{\mathcal{H}} a = \mathcal H_0 a + \iu\diag(\eta_j) a.
  \end{align}
  \item Model the diagonal part with local dissipation (observed: $\kappa_j$, not observed: $\eta_j^l$) and amplification ($\eta_j^g$) which introduces quantum noise:
  \begin{align}
    \iu\partial_t a
    &= \mathcal H_0 a + \iu\diag(\eta_j) a + \iu F = \tilde{\mathcal{H}} a + \iu F, \\
    \eta_j
    &= -\kappa_j - \eta_j^l + \eta_j^g, \\
    F_j
    &= \sqrt{2\kappa_j} O_{j, \text{in}} + \sqrt{2\eta_j^l}L_{j, \text{in}} - \sqrt{2\eta_j^g} G^\dagger_{j, \text{in}}
  \end{align}
  with $\kappa_j, \eta_j^l, \eta_j^g \geq 0$.
  For the observed channels, the Input-Output-relation is then given by
  \begin{align}
    O_{j, \text{out}} = O_{j, \text{in}} - \sqrt{2\kappa_j} a_j.
  \end{align}
\end{enumerate}

The resulting equations describe a weakly interacting multi-mode system in a fully quantum-mechanical setting which reproduces up to a unitary basis-transformation the non-Hermitian Hamiltonian $H$ in the classical limit of neglecting the input noise $F$.

\section{QUANTOS model}
Finally, we design our coupled mode system such that the corresponding non-Hermitian Hamiltonian $H$ enters a NH topological phase, as classically studied in  \cite{Budich2020}.
Concretely, our setup is described by the non-Hermitian Bloch-Hamiltonian
\begin{align}
  \tilde H(k) = (t_1 + t_2 \cos k, 0, t_2 \sin k - \iu \gamma) \cdot \bm{\sigma} \label{eq:Bloch_NTOS}
\end{align}
where two neighboring modes form a unit-cell of a lattice translation-invariant system, and the vector of standard Pauli matrices $\bm{\sigma}$ acts within the two-mode system formed by each unit cell.
The NH topological phase of this model is characterized by the spectral winding number
\begin{align}
  \nu = \frac{1}{2\pi \iu} \oint_0^{2\pi} \dd{k} \left\{\partial_k
  \ln[\det(\tilde H(k))]\right\}
\end{align}
which acquires a non-vanishing value if $||t_1| - |t_2|| < |\gamma| < ||t_1| + |t_2||$.

Switching to a real space description of a system with $N=2p-1$ modes in a broken ring geometry, where the ends are weakly coupled with a parameter $\Gamma$, the NH Hamiltonian matrix reads as
\begin{align}
  \tilde{\mathcal H} =
  \begin{pmatrix}
    -\iu\gamma & t_1 & \frac{\iu t_2}{2} & \frac{t_2}{2} & 0 & \ldots & \Gamma \\
    t_1 & \iu\gamma & \frac{t_2}{2} & \frac{-\iu t_2}{2} & 0 & \ldots & 0 \\
    \frac{-\iu t_2}{2} & \frac{t_2}{2} & -\iu\gamma & t_1 & \frac{\iu t_2}{2} & \ldots & 0 \\
    \frac{t_2}{2} & \frac{\iu t_2}{2} & t_1 & \iu\gamma & \ddots & \ddots & 0 \\
    0 & 0 & \frac{-\iu t_2}{2} & \ddots & \ddots & t_1 & \frac{\iu t_2}{2} \\
    \vdots & \vdots & \vdots & \ddots & t_1 & \iu\gamma & \frac{t_2}{2} \\
    \Gamma & 0 & 0 & \ldots & \frac{-\iu t_2}{2} & \frac{t_2}{2} & -\iu\gamma
  \end{pmatrix}.
\end{align}

The Hermitian part corresponds to a coherent mode coupling between nearest-neighboring unit cells.
The anti-Hermitian part is already diagonal in the mode basis, which corresponds to the above local dissipation assumption:
\begin{align}
  \mathcal{H}_\text{AH} = \iu\gamma\diag(-1, 1, -1, \ldots, 1, -1) \label{eq:H_diag}
\end{align}
and thus we can easily translate this into our generic quantum theoretical setting outlined above.
Concretely, the modes with $+\iu\gamma$ are coupled to gain channels ($\eta_j^g$) and the $-\iu\gamma$ to loss channels ($\eta_j^l$).
Furthermore, the first and the last mode are assumed to be observed.
That way, we arrive at the QUANTOS model described in the main text.

\section{Robustness against imbalance of gain and loss}
\label{app:balanced}
The balanced gain and loss in our model system may remind the reader of local $PT$-symmetry which naturally raises the question if our findings, in particular the exponential amplification of the Fisher information with system size, rely on the presence of such a symmetry.
To this end, we study an unbalanced QUANTOS model which is described by the non-Hermitian Bloch-Hamiltonian (cf.~Eq.~(\ref{eq:Bloch_NTOS}))
\begin{align}
  \tilde H^\text{S}(k) = \tilde H(k) - \iu\epsilon\gamma\sigma_0 \label{eq:H_shifted}
\end{align}
with $\epsilon \geq 0$ which gives (cf.~Eq.~(\ref{eq:H_diag}))
\begin{align}
  \mathcal H_\text{AH}^\text{S} = \iu\gamma\diag\left(-(1+\epsilon), 1-\epsilon, -(1+\epsilon), \ldots\right).
\end{align}

\begin{figure}
  \includegraphics{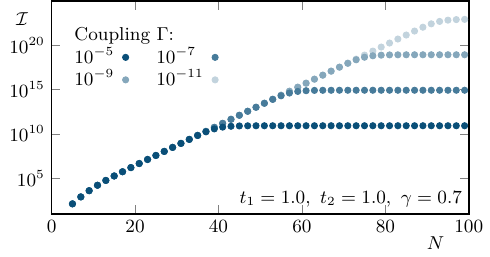}
  \caption{
    Scaling with the system size $N$ of the Fisher information $\mathcal I$ for several values of $\Gamma$ of the unbalanced QUANTOS model (cf.~Eq.~(\ref{eq:H_shifted})).
    In comparison with Fig.~\ref{fig:exp_amp_region}(a) we can see that the exponential growth rate $\alpha$ is smaller and thus, more sites $N$ are needed until the Fisher information $\mathcal I$ reaches saturation.
    The parameters were chosen $t_1 = t_2 = 1,\ \gamma = 0.7$ and $\epsilon = 0.1$.
  }
  \label{fig:shifted_Fisher}
\end{figure}

We found that breaking the symmetry by unbalancing gain and loss does not destroy the exponential enhancement (cf.~Fig.~\ref{fig:shifted_Fisher}) as long as the gain exceeds the additional loss (i.e.~$0 \leq \epsilon < 1$) but only alters the exponential growth rate $\alpha$ (see Eq.~(\ref{eqn:one})).
Particularly, $\alpha$ decreases with larger value of $\epsilon$ until $\epsilon = 1$ where gain and additional loss become equal which results in $\alpha = 0$.
Thus, more sites $N$ are needed in order to reach the saturation value.
With this, we can conclude that the $PT$-symmetry is not necessary for the working principle of the sensor.

\end{document}